\documentclass[showpacs,twocolumn,floatfix,amsmath,amssymb,superscriptaddress,prl]{revtex4}
\usepackage{epsfig}
\usepackage{graphicx}
\usepackage{longtable}
\usepackage{color}

\begin{document}

\title{Edge states and topological insulating phases  generated by  curving a nanowire with Rashba spin-orbit coupling}

\author{Paola Gentile}
\affiliation{CNR-SPIN and Dipartimento di Fisica ``E. R. Caianiello'',
Universit\`a di Salerno I-84084 Fisciano (Salerno), Italy
, I-84084 Fisciano (Salerno), Italy}

\author{Mario Cuoco}
\affiliation{CNR-SPIN and Dipartimento di Fisica ``E. R. Caianiello'',
Universit\`a di Salerno I-84084 Fisciano (Salerno), Italy, I-84084 Fisciano (Salerno), Italy}

\author{Carmine Ortix}
\affiliation{Institute for Theoretical Solid State Physics, IFW-Dresden, Helmholtzstr. 20, D-01069 Dresden, Germany}
\affiliation{Institute for Theoretical Physics, Utrecht University, Leuvenlaan 4, 3584 CE Utrecht, The Netherlands}

\date{\today}

\begin{abstract}
We prove that curvature effects in low-dimensional nanomaterials can promote the generation of topological states of matter by considering the paradigmatic example of  quantum wires with Rashba spin-orbit coupling,  which are periodically corrugated at the nanometer scale. 
The effect of the periodic curvature generally results in the appearance of insulating phases with a corresponding  novel butterfly spectrum characterized by the formation of fine measure complex regions of forbidden energies. When the Fermi energy lies in the gaps, the system displays  localized end states protected by topology.
We further show that for certain corrugation periods the system possesses topologically non-trivial insulating  phases at half-filling. Our results suggest that the local curvature and the topology of the electronic states are inextricably intertwined in geometrically deformed nanomaterials. 
\end{abstract}

\pacs{73.63.Nm, 73.21.Cd, 03.65.Vf, 73.43.-f}

\maketitle

\paragraph {Introduction -- } 
In recent years, topological non-trivial states of matter have been the subject of growing interest \cite{has10,moo10,qi11}. Topologically non-trivial electronic phases were discovered in time-reversal invariant insulators -- leading to the quantum spin Hall (QSH) effect in two-dimensional systems \cite{kan05b,kan05,ber06}, and to the existence of protected two-dimensional Dirac cones on the surface of three-dimensional topological insulators \cite{fu07b,hsi08,zha09,che09}--  as well as in insulators with additional specific  crystal point group symmetries \cite{fu11}.  Likewise, topological states in superconducting systems have been intensively studied since their midgap excitations can be potentially used to encode quantum bits with an unusually long coherence time \cite{ali12,bee13}.  
Among the different solid-state platforms where topological quantum states of matter are expected to arise, low-dimensional semiconductor nanomaterials play undoubtedly a primary role. 
The QSH effect, for instance, was first theoretically predicted \cite{ber06} and later experimentally proved in HgTe quantum wells \cite{kon07}. An heterostructure comprising a semiconductor nanowire with strong spin-orbit coupling and a conventional s-wave superconductor has been suggested to host a  topological superconducting phase \cite{lut10,ore10} with signatures of midgap  Majorana bound states already reported \cite{mou12}.

Apart from these conventional material geometries, rapid advances in nanostructuring techniques have enabled the synthesis of novel low-dimensional nanostructures  in which flexible semiconductor nanomaterials can be bent into curved, deformable objects such as spiral-like nanotubes \cite{pri00,sch01}, nanohelices \cite{zha02} and even complex  nanoarchitectures  resembling structures that form naturally in the most basic forms of life \cite{xu15}. On the one hand,  these next-generation nanomaterials carry an enormous potential in electronics, ranging from flexible displays to the integration of semiconductor electronics with the soft, curvilinear surfaces of the human body \cite{rog11,jo15}. 
On the other hand, the very fundamental quantum mechanical properties of the charge carriers in these nanomaterials are strongly affected by the curved background in which they live \cite{dew57}. Consequently, unique curvature-induced electronic and transport properties have been identified. These include, but are not limited to,  the appearance of winding-generated bound states \cite{ort10},  
and an extremely large anisotropic magnetoresistance in non-magnetic and spin-orbit-free semiconducting rolled-up nanotubes \cite{cha14}.

\begin{figure}
\includegraphics[width=\columnwidth]{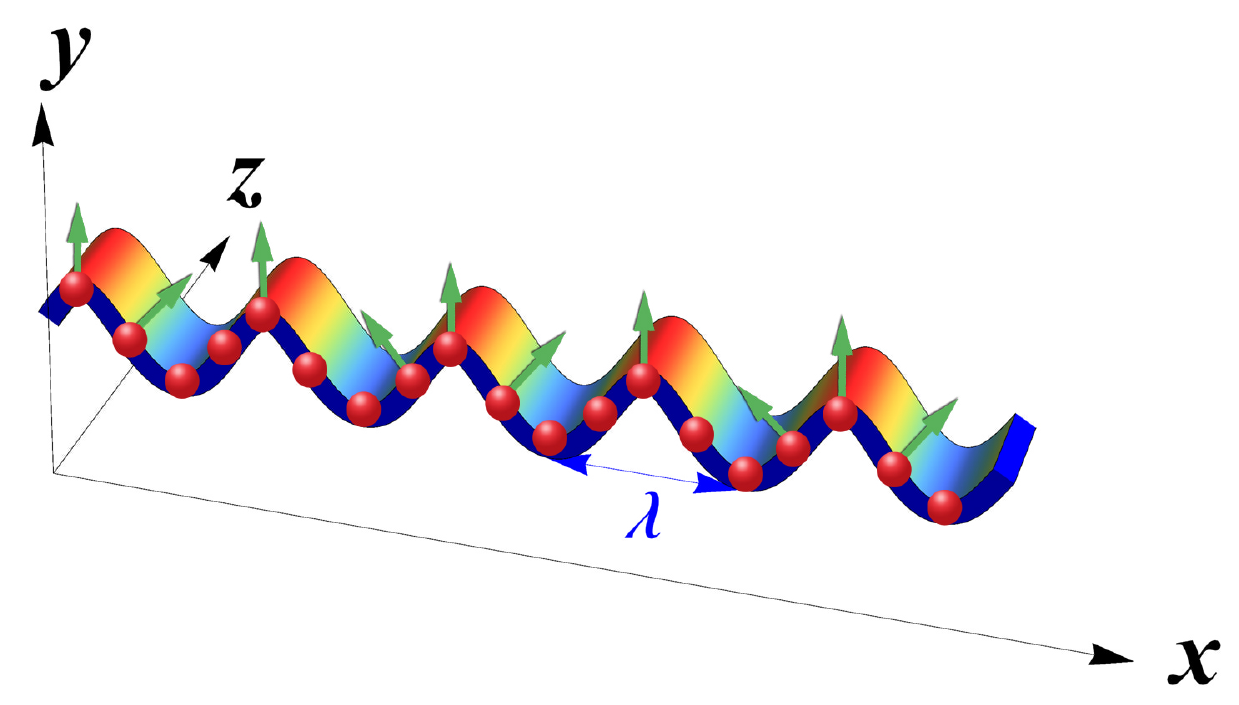}
\caption{(color online) Schematic view of a planar nanocorrugated nanowire with corrugation period $\lambda$. The red spheres indicate atomic sites while the green arrows are the corresponding local directions of the Rashba spin-orbit field axis. }
\label{fig:fig1}
\end{figure}

A relevant question that naturally arises is whether and under what circumstances the interplay between curvature effects on the electronic properties and the topology of the ground state of a low-dimensional solid-state system can be significant. In this Letter, we show that not only such interplay is relevant but also that the curved geometry of a  bent
nanomaterial can promote the generation of non-trivial edge states and topological insulating phases. 
We prove the assertion above by considering the simple example of a nanowire with Rashba spin-orbit coupling which is corrugated to acquire a "serpentine"-like planar shape [c.f. Fig.~\ref{fig:fig1}].  The ensuing periodic canting of the spin-orbit field axis leads to a miniband structure and a correspondent metal-insulator transition at different filling fractions. 
This, in turn, leads to a butterfly spectrum that, however, is neither of the Hofstadter type \cite{hof76} nor of the Moth type \cite{ost05}. A subsequent analysis of the system with open boundary conditions shows the occurrence of midgap edge states, which we show to be topological in nature.
Finally, for certain corrugation periods, the system can display topologically non-trivial insulating phases at half-filling. 

\paragraph{Metal-insulator transition  -- } Our starting point is an effective continuum ${\bf k \cdot p}$ model for a planarly curved quantum wire with a Rashba spin-orbit term originating from an electric field pointing in the direction perpendicular to the plane of the wire (the ${\hat z}$ axis in Fig.~\ref{fig:fig1}). In its symmetric, Hermitian, form it reads:
\begin{equation}
{\cal H}_{\bf k \cdot p}= -\dfrac{\hbar^2}{2 m^{\star}} \partial_s^2 - \dfrac{i \alpha_R}{2} \left[\tau_N(s) \partial_s + \partial_s \tau_N(s)\right], 
\label{eq:hamiltoniankp}
\end{equation}
where $s$ is the arclength of the bent nanowire measured from a reference point, $m^{\star}$ is the effective mass of the charge carriers, and  $\alpha_R$ is the strength of the Rashba spin-orbit coupling. Finally $\tau_N(s)$ is a local Pauli matrix comoving with the electrons as they propagate along $s$, and explicitly reads $\tau_N(s) = \boldsymbol{\tau} \cdot \hat{\cal N}(s)$, where $\hat{\cal N}(s)$ is the local normal direction of the curved wire, while the $\boldsymbol{\tau}$'s are the usual Pauli matrices. The Hamiltonian in Eq.~\ref{eq:hamiltoniankp} corresponds to the effective continuum model for curved quantum wires of Ref.~\onlinecite{gen13} in the absence of strain-induced effects, and can be mapped onto the model for a conventional quantum wire with a locally varying spin-orbit field axis [c.f. Fig.~\ref{fig:fig1}]. The latter can be determined by writing the normal direction in the Euclidean space in terms of an angle $\theta(s)$ as $\hat{\cal N}(s) = \left\{ \sin{\theta(s)} , \cos{\theta(s)} ,0 \right\}$, and using the Frenet-Serret type equation of motion $\partial_s \hat{\cal N}(s) = - \kappa(s) \hat{\cal T}(s)$ with $\hat{\cal T}(s) = \left\{\cos{\theta(s)} , -\sin{\theta(s)},0 \right\}$ the tangential direction and $\kappa(s)$ the local curvature. It then follows that the local direction of the spin-orbit field axis is entirely determined by the curvature of the quantum wire via $\theta(s) = - \int^s \kappa(s^{\prime}) d s^{\prime}$. For a periodically corrugated quantum wire, it also implies  that the spin-orbit field axis undergoes a periodic canting with the maximum canting angle proportional to the curvature. 

In order to study the effect of such a periodic canting of the spin-orbit field axis on the electronic properties of a quantum wire, we next introduce a tight-binding model obtained by discretizing Eq.~\ref{eq:hamiltoniankp} on a lattice. It can be written as: 
\begin{equation}
{\cal H}=\sum_{j} \sum_{\sigma,\sigma'=\uparrow,\downarrow}
c^{\dag}_{j,\sigma}(t \ \delta_{\sigma, \sigma'} +  \hat{\alpha}_{j,j+1}^{\sigma,\sigma'})c_{j+1,\sigma'} +{\it h.c} , 
\label{eq:hamiltonian}
\end{equation}
 where $c_{j,\sigma}^{\dag}, c_{j,\sigma}$ are operators creating and annihilating, respectively, an electron at the $j$-th site with spin projection $\sigma= \uparrow, \downarrow$ along the $z-$axis, $t$ is the hopping amplitude between nearest-neighbor sites,  and the spin-dependent nearest-neighbor hopping amplitudes are 
 \begin{equation}
\hat{\alpha}_{j,j+1}=  i \alpha_R \, \left[ \tau_x \, g_j^{x} + \, \tau_y \, g_j^y \right]. 
\end{equation}
 In the equation above, $g_j^{x}= \sin{\theta(s_j)} + \sin{\theta(s_{j+1})}$ and $g_j^{y}= \cos{\theta(s_j)} + \cos{\theta(s_{j+1})}$, which are determined by the position of the atoms along the quantum wire and the specific geometric shape of the nanocorrugation. For the latter, we assume a simple sinusoidal form with parametric equation in Euclidean space ${\bf r}=\left\{x, A \sin{ (2 \pi x) / \lambda}, 0  \right\}$ with corrugation period $\lambda$ and corrugation height $A$. The atomic positions can be instead written  as 
 $s_j / \lambda= p j / q + \varphi / (2 \pi)$, where $p$ and $q$ are integers whose ratio $p/q = a / \lambda$ ($a$ being the lattice constant) while $\phi \in \left[0 , 1\right]$ accounts for non-equivalent displacements of the atoms in one corrugation period. As a result, the local spin-dependent hopping amplitudes render a one-dimensional superlattice with superlattice constant $p \lambda$. 
 Although $p/q \ll 1$ in a quantum wire smoothly corrugated at the tens of nanometers scale, we will extend our analysis to the full range $0<p/q<1$ taking Eq.~\ref{eq:hamiltonian} as a general model system to study curvature effects even  in microscopically buckled atomic chains.
 
\begin{figure}
\includegraphics[width=.9\columnwidth]{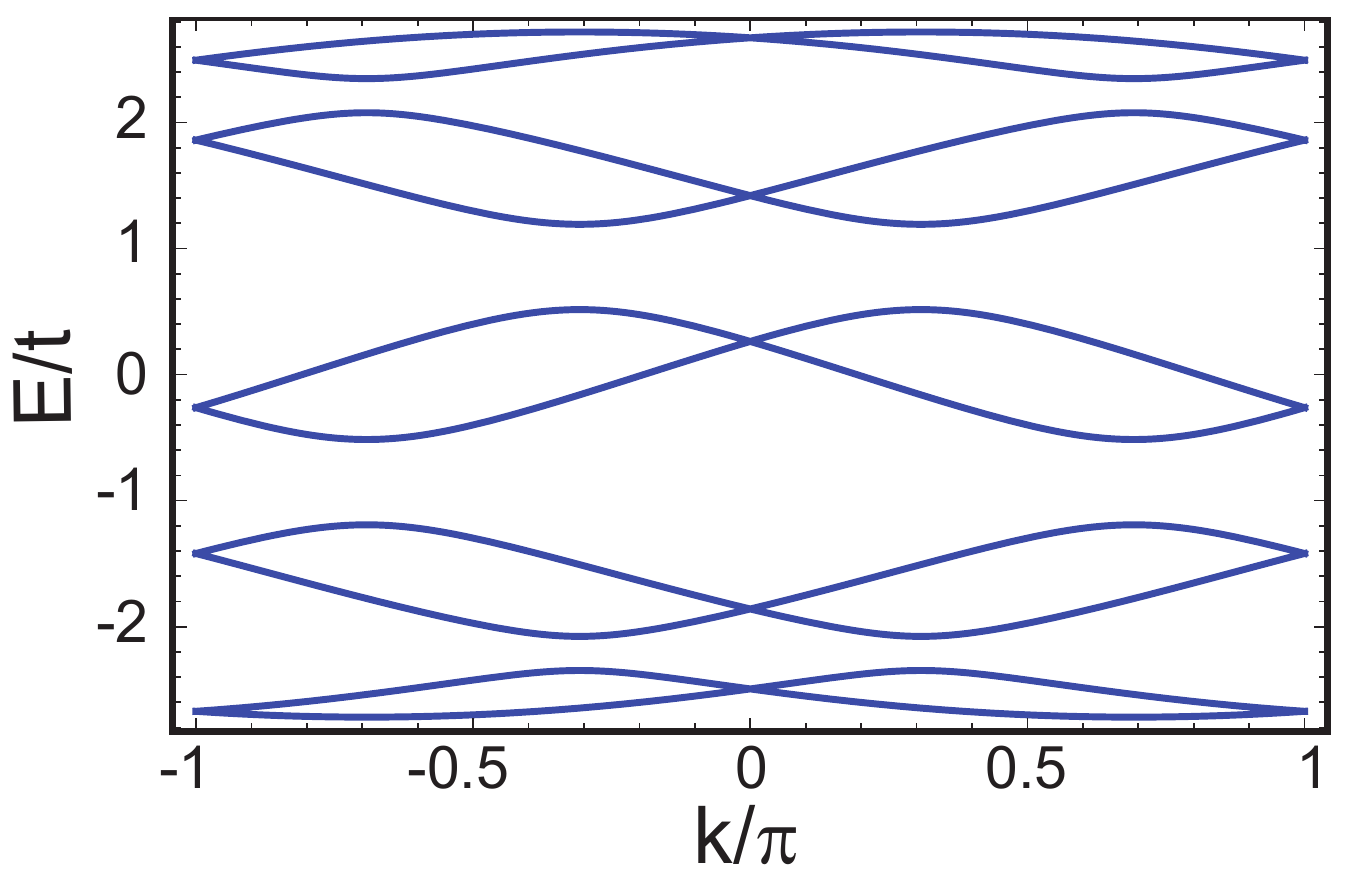}
\caption{(color online) Band structure of a nanocorrugated wire with Rashba spin-orbit coupling in the first mBZ for  $p/q=1/5$, $\alpha_R = 0.6 t$, $A/\lambda=1$, and  
the displacement field $\phi=0$. 
Momenta are measured in units of $1 / \lambda$. The canting of the spin-orbit field axis leads to bandgap openings at unpinned momenta in the mBZ.}
\label{fig:fig2}
\end{figure} 

Fig.~\ref{fig:fig2} shows the band structure for $p/q=1/5$ in the first mini-Brillouin zone (mBZ) $k \in \left[ - \pi /( p  \lambda) , \pi /(p  \lambda) \right]$. The presence of the spin-orbit interaction terms removes   the spin degeneracy of the minibands except at the time-reversal ${\cal T}$  invariant points $k=0,  \pi / ( p  \lambda) $ where the eigenstates are Kramers degenerate. This, however, does not preclude {\it a priori} band gaps opening at different points in the mBZ. And indeed Fig.~\ref{fig:fig2} shows the occurrence of full bandgaps at the filling fractions $\nu = n/ q $, with $n$ integer, whose magnitude monotonically increases with the corrugation height $A$. 
We find the opening of bandgaps for finite values of  the corrugation height $A$ to occur  for $p/q=1/6,1/7$ [see Supplemental Material]. 
 Henceforth, the periodic corrugation of the nanowire generally induces a metal-insulator transition, and thus defines a nanoflex transistor switch, the status of which is ``on''   when the nanowire is flat and ``off"  when the nanowire is planarly curved. A similar nanoscale transistor switch has been suggested in nanohelices without spin-orbit coupling but subject to an  externally applied rotating electric field \cite{qi09}. 

\begin{figure}[t!]
\includegraphics[width=.85\columnwidth]{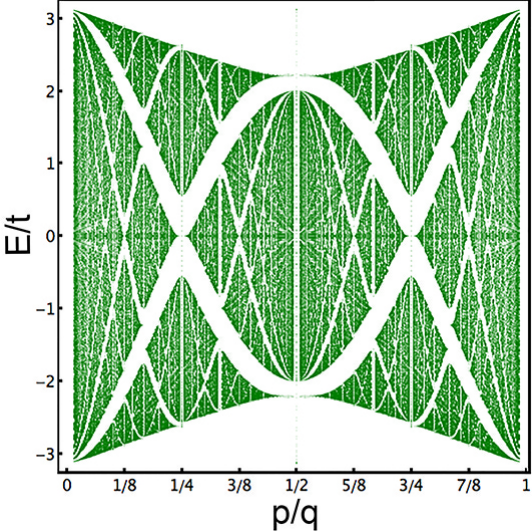}
\caption{(color online) Butterfly spectrum for the superlattice system of Eq.~\ref{eq:hamiltonian} as a function of $p/q$. We have set the corrugation height $A= 5 \lambda$. The spectrum clearly exhibits a self-similar structure.}
\label{fig:fig3}
\end{figure} 

\paragraph{Butterfly spectrum and topological edge states -- }
To further verify that the metal-insulator transition is a generic feature due to the periodic canting of the Rashba spin-orbit field axis, 
we have diagonalized the Hamiltonian Eq.~\ref{eq:hamiltonian}  for different commensurate superlattices, changing the displacement field $\phi$ in the 
$\left [0 , 1 \right]$ interval while keeping the ratio $A / \lambda$ fixed. 
Fig.~\ref{fig:fig3}  shows the ensuing energy spectrum as a function of the ratio between the atomic lattice constant and the corrugation period. 
It exhibits two main holes of forbidden energies with a double wing shape joining at $p/q=1/4$ and $p/q=3/4$. The generation of replicas of such twin wings of holes of finite measure and various sizes, the largest of which join at $p/q= (2 n +1 )/8$, evidences a self-similar ``butterfly"-like structure of the electronic spectrum. 
Although a rigorous proof cannot be made, this also reminds one of a fractal structure. Furthermore, the present spectrum is manifestly different in shape both from the Hofstadter butterfly realized  in conventional two-dimensional electron gases \cite{hof76}, graphene moir\'e superlattices \cite{pon13} and one-dimensional optical superlattices \cite{aid13,miy13}, as well as from the moth butterfly discussed in the context of cold atoms in non-Abelian gauge potentials \cite{ost05}. 

\begin{figure}
\includegraphics[width=.85\columnwidth]{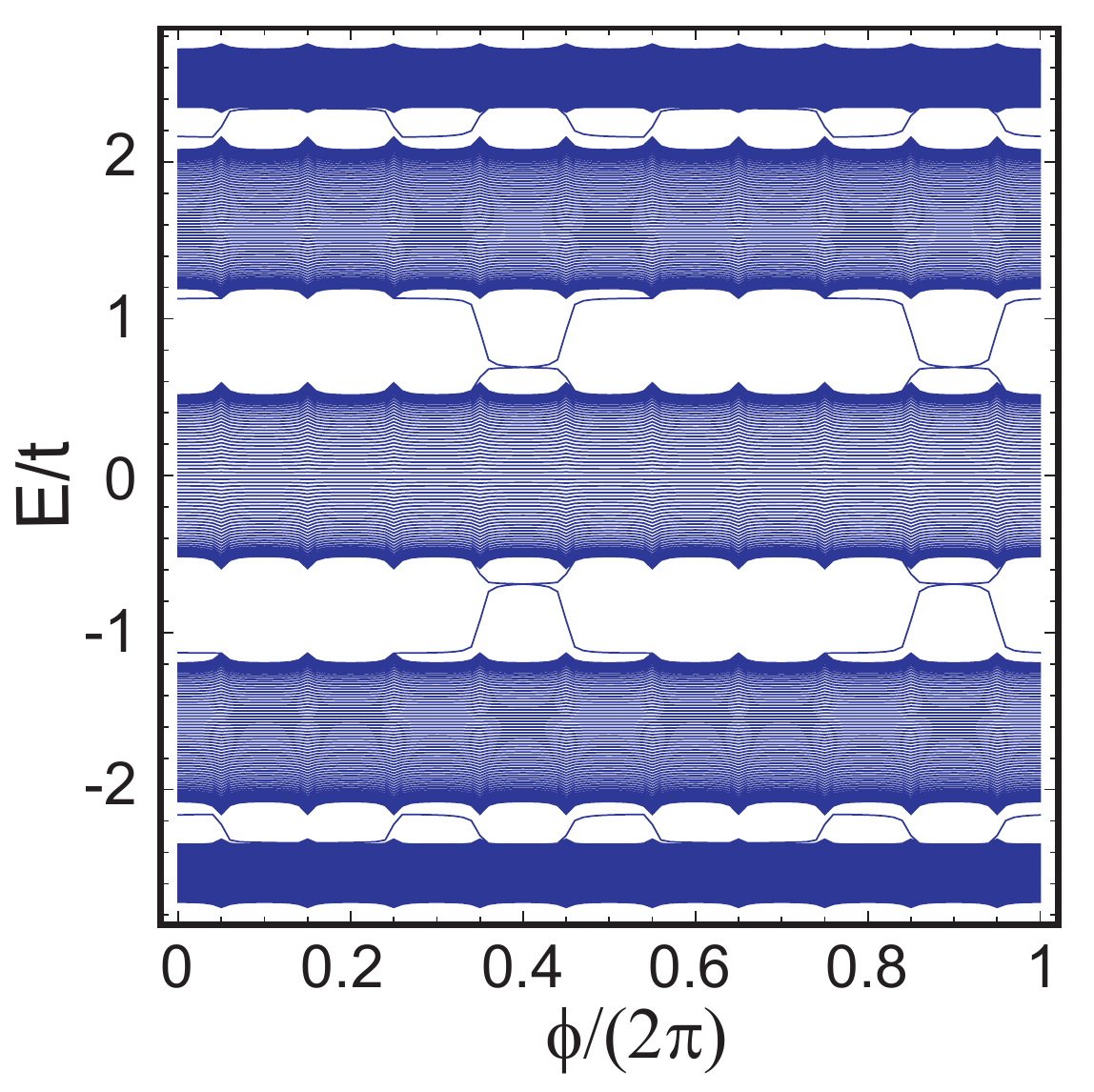}
\caption{(color online) Energy spectrum of the model Hamiltonian Eq.~\ref{eq:hamiltonian} for $p/q=1/5$, $\alpha_R=0.6$ and $A/\lambda=1$. The spectrum has been obtained by exact diagonalization of the Hamiltonian for a finite atomic chain containing $N=650$ atoms and open boundary conditions.}
\label{fig:fig4}
\end{figure}

Having established the ubiquitous presence of insulating phases in our nanocorrugated nanowire, we have then analyzed the possible occurrence of localized edge states within the curvature-induced gaps while sweeping the displacement field $\phi$, as routinely done in one-dimensional optical superlattices with an additional dynamical parameter \cite{lan12,mar15}. Fig.~\ref{fig:fig4} shows the energy spectrum for a finite size system with open boundary conditions considering the superlattice periods $\lambda=5 a$. 
Within all gaps, there appear edge states  localized at the left and right boundaries of the atomic chain whose energy changes continuously with  $\phi$,  connecting the valence to the conduction bulk bands. Since the appearance of such edge states can be generally attributed to non-trivial topological properties of bulk systems, we have then analyzed the topological character of the ground state of the system with the displacement field $\phi$ playing the role of an artificial dimension \cite{zha13}. It can be shown that the addition of this extra dimension breaks the ``time-reversal" symmetry constraint $\Theta^{-1} {\cal H}(k , \phi) \Theta \neq {\cal H}(-k , -\phi)$, thereby allowing for insulating phases with non-zero Chern numbers. We have therefore computed the Chern numbers in the insulating phases of Fig.~\ref{fig:fig2} using the method outlined in Ref.~\onlinecite{tak05} and have found Chern number $C=\pm 4$ and $C=\pm 8$ for the two central and two exterior gaps respectively. This result is in perfect agreement with the foregoing analysis of the spectrum with open boundary conditions, which, depending on the filling fraction, indeed shows four or eight edge states per edge composed of the two spin species [c.f. Fig.~\ref{fig:fig3}].  We have 
verified that in-gap edge states protected by a non-zero Chern number also appear at different values of $p/q$ [see the Supplemental Material], thereby directly proving the general occurrence of edge states topological in nature in a nanocorrugated quantum wire.

\begin{figure}
\includegraphics[width=.9\columnwidth]{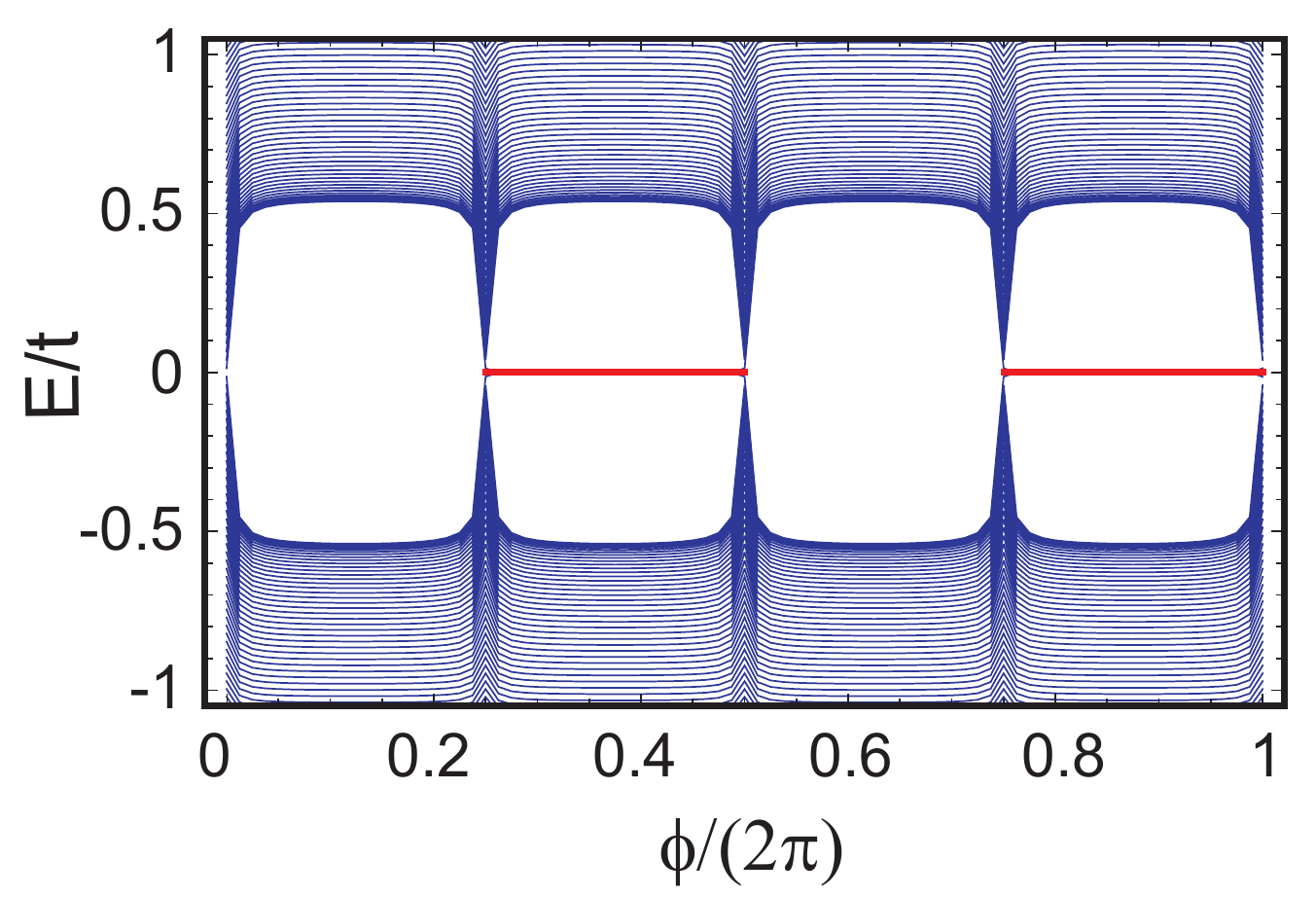}
\caption{(color online) Energy spectrum of the model Hamiltonian Eq.~\ref{eq:hamiltonian} for $p/q=1/4$, $\alpha_R=0.6$ and $A/\lambda=1$. The spectrum has been obtained by exact diagonalization of the Hamiltonian for a finite atomic chain containing $N=640$ atoms and open boundary conditions.}
\label{fig:fig5}
\end{figure} 

\paragraph{Topological insulating phases -- }  We finally show that a periodic canting of the Rasha spin-orbit field axis can also lead to one-dimensional topological insulating phases. This occurs at the central points of the wings in our butterfly spectrum of Fig.~\ref{fig:fig3}. 
For all values of $\phi$, the system possesses time-reversal symmetry $\Theta^{-1} {\cal H}(k , \phi) \Theta  =  {\cal H}(-k ,\phi)$ with $\Theta^2=-1$ as required for spin-one-half  fermions, a unitary chiral symmetry ${\cal C}^{-1} {\cal H}(k, \phi) {\cal C}= - {\cal H}(k, \phi)$, and an antiunitary symplectic particle-hole symmetry ${\cal P}^{-1}{\cal H}(k, \phi) {\cal P}= -{\cal H}(-k , \phi)$ with ${\cal P}^2 = -1$. This also implies that at these selected $p/q$ values, the system  is in the chiral symplectic symmetry class CII of the Altland-Zirnbauer classification \cite{alt97}, and thus possesses an integer ${\mathcal Z}$ topological invariant provided it is insulating at half-filling. In terms of the displacement field $\phi$, we find that the half-filling gap has an even $q$ number of closing-reopening points [c.f. Fig.~\ref{fig:fig5} and Supplemental Material], which is an unambiguous signal of the occurrence of  a topological phase transition between insulating phases with different ${\mathcal Z}$ invariant. To verify this point, we have thus considered a superlattice system with $\lambda = 4 a$ and open boundary conditions. The ensuing energy spectrum as a function of the displacement field $\phi$ is shown in Fig.~\ref{fig:fig5}. 
It shows the presence of two spin-degenerate topologically protected zero-energy modes between the gap closing-reopening points, thereby explicitly proving that the geometric displacement of the atoms in a nanocorrugated nanowire drives a  topological phase transition between a ${\mathcal Z}=2$  and  a ${\mathcal Z}=0$ insulating state. 
 Precisely the same feature is encountered at the central points of the wings replicas in the butterfly spectrum of Fig.~\ref{fig:fig3} [see the Supplemental Material].

\paragraph{Conclusions -- } To sum up, we have analyzed curvature effects on the electronic states of a nanocorrugated quantum wire with Rashba spin-orbit interaction. We have shown that the ensuing periodic canting of the Rashba spin-orbit field yields a novel butterfly spectrum with insulating phases characterized by the  presence of edge states, which are topological in origin. 
We have also shown that the curved geometry of such a wrinkled nanowire can promote the onset of  one-dimensional topological insulating phases with topologically protected zero end modes. 
Our results constitute a proof-of-principle demonstration that the geometric curvature of  flexible nanomaterials can impact the topology of the electronic states in low-dimensional systems, and can thus  inaugurate the search for other topological non-trivial states of matter in solid-state systems manufactured with the most advanced nanotechnology techniques.

\paragraph{Acknowledgements --} We acknowledge the financial support of the Future and Emerging Technologies (FET) programme within the Seventh Framework Programme for Research of the European Commission, under FET-Open grant number: 618083 (CNTQC).  CO thanks the Deutsche Forschungsgemeinschaft (grant No. OR 404/1-1) for support.

\end{document}